# Digitizing analogic spectrograms recorded by the Nançay Decameter Array on 35 mm film rolls from 1970 to 1990

Baptiste Cecconi[1,2], Laurent Lamy[1,2], Laurent Denis[2], Philippe Zarka[1,2], Agnès Fave[1], Marie-Pierre Issartel[1], Marie-Agnès Dubos[3] Corentin Louis[1], Pierre Le Sidaner[4] and Véronique Stoll[3]

[1]LESIA, Observatoire de Paris, CNRS, PSL, Sorbonne Université, Meudon, France, [2]Station de Radioastronomie de Nançay, Observatoire de Paris, CNRS, PSL, Université d'Orléans, Nançay, France, [3]Bibliothèque, Observatoire de Paris, CNRS, PSL, Paris, France, [4]DIO, Observatoire de Paris, CNRS, PSL, Paris, France.
Corresponding author: Baptiste Cecconi (baptiste.cecconi@observatoiredeparis.psl.eu)

The Nançay Decameter Array (NDA), which has now passed 40 years old, acquires daily observations of Jovian and Solar low frequency radio emissions over a continuous spectrum ranging from 10 up to 100MHz, forming the largest database of LW radio observations of these two bodies. It also intermittently observed intense radio sources since its opening in 1977. Before that date, decametric observations were conducted on the same site with an interferometer formed of a pair of log-periodic Yagi antennas mounted on mobile booms. These observations have been recorded with a series of analogic recorders (before 1990) and then digital receivers (after 1990), with increasing performances and sensitivities.

The NDA scientific team recently retrieved and inventoried the archives of analogic data (35mm film rolls) covering two decades (1970 to 1990). We now plan to digitize those observations, in order to recover their scientific value and to include them into the currently operational database covering a time span starting in 1990 up to now, still adding new files every day. This modern and interoperable database has virtual observatory interfaces. It is a required element to foster scientific data exploitation, including Jovian and Solar data analysis over long timescales. We present the status of this project.

## 1. Introduction

The Nançay Decameter Array (NDA), hosted at the *Station de Radioastronomie de Nançay* (Sologne, France), is observing quasi-continuously low frequency radio emissions of the Solar corona and Jupiter (and occasionally other intense radio sources) in the spectral range 10-100 MHz since 1977 [1, 2, 3]. During the preceding decade, decametric observations were already carried on with an interferometer (hereafter referred to as the Nançay Decameter Interferometer, NDI) composed of a pair of log-periodic Yagi antenna mounted on mobile booms [4], whose receivers were originally tested at the Arecibo Observatory [5]. These observations have been acquired with a series of locally developed analogic (NDI/NDA before 1990) and digital (NDA after 1990) receivers with increasing performances. The NDA scientific team recently retrieved the archive of analogic data recorded on a series of 35mm 100ft film rolls covering two decades (1970 to 1990) of observations. We now plan to digitize this data collection in order to favor its scientific exploitation and extend the current NDA database, which already contains all digital data recorded since September 1990. This database is updated daily with new observations. It has been recently reorganized and it now implements modern interoperable interfaces (e.g., virtual observatory standards) [3]. This database will ultimately host the historical digitized data, providing thus a unique data collection spanning on more than 4 decades, which is more than 3 solar cycles and 3 Jovian revolutions around the Sun.

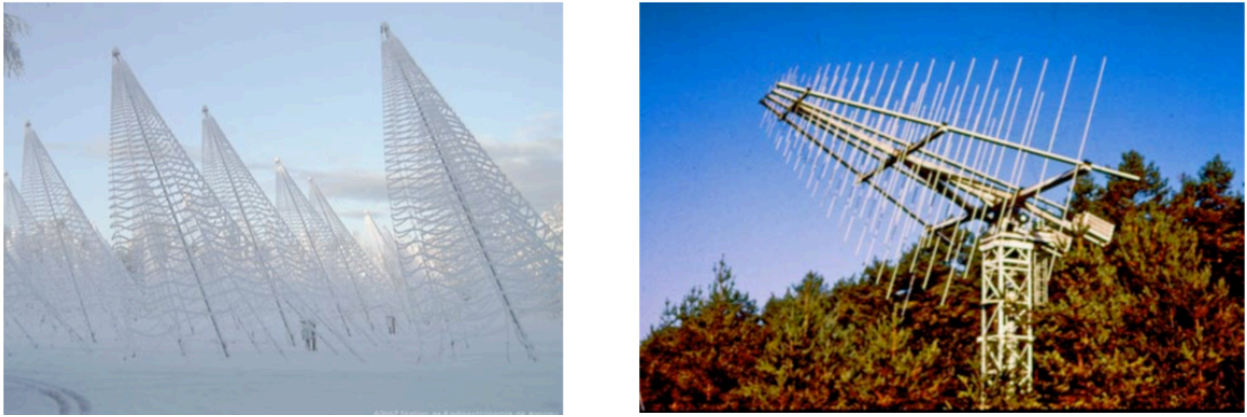

*Figure 1. (Left) The Nançay Decameter Array consists in 144 helicoidal antennae divided in two left-handed and right-handed polarized subsets [1]. (Right) Historic log-periodic Yagi antenna in Nançay [2].*

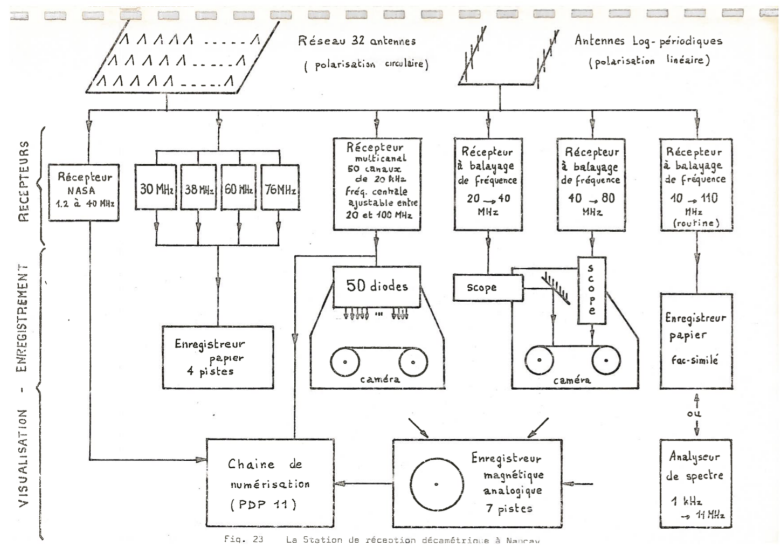

*Figure 2. Instrumental setup of NDA and Yagi antennae during the Voyager flybys of Jupiter [3].*

## 2. Film roll collection

The analogic data collection is composed of 1492 films, which are standard 35mm 100ft rolls. The data was projected in real-time on the roll, with a continuous sliding recording device, usually running at a speed of 3 cm per minutes. Each film thus consists in ~30 meter-length spectrograms recorded continuously on the film roll and generally includes several observing sequences of the Sun and Jupiter. The rolls have been stored in their original metallic canister after chemical development. Most of the film rolls are in healthy conditions, except for a limited series, which have been put in contact with water. This small set shows rusted canisters and glued or damaged photographic gelatin. Most of rolls are numbered with stickers on the film canister. These manual annotations include the name of the receiver, the date(s) of observations and sometimes the observed radio source. Other annotations are reported on the film itself, mostly with hand writing. The film rolls contain time stamps (day of year, hour of day, and decimal number of minutes) and temporal tick marks in the form of 2 dashed lines: a high-resolution line with 1 second consecutive plain and empty segment of equal lengths (30 pairs of plain and empty segments make 1 minutes of data), and a low-resolution line with 1 minute consecutive plain and empty segments of equal lengths (5 pairs of plain and empty segments make 10 minutes of data, i.e., the interval between 2 successive time stamps.

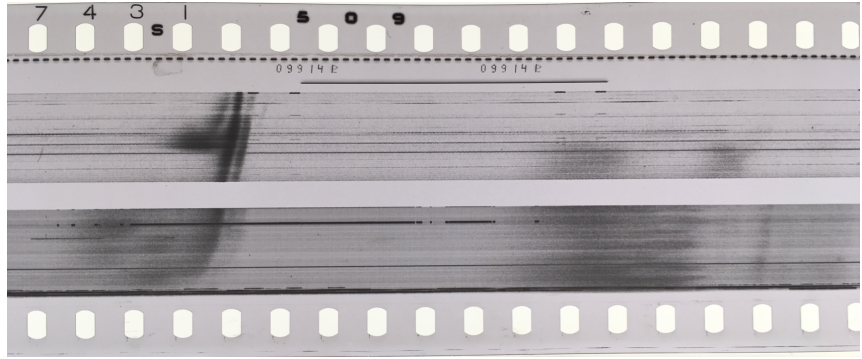

*Figure 3. Example of film roll, showing time stamps (day 099 of current year, 14h, start of 10th minute, and long/short 1-minute/1-second ticks. The spectrogram display a solar Type III burst.*

Some rolls contain data and annotation recorded on the side perforations, usually due to a recorder misalignment.
Most the data collection contains recording of Sun and Jupiter observations. However, a few early rolls correspond to astrophysical objects observations. A noticeable series concerns solar and pulsar observations led at Arecibo in July and August 1970 [5], including test measurements of the first identified pulsar CP 1919 ([PSR B1919+21](#)) discovered by Jocelyn Bell in 1967.

An exhaustive inventory has been built, listing all the metadata available by visual inspection of the canister and roll (by without unrolling it). Precisely, the inventory file so far contains, for each roll: the name of the person filling the file, the date of reporting, the roll number, description and time interval (as written on the canister), its physical location, its preservation state, other observations if any. It will be completed at the digitization step by additional metadata on the roll state and on the carried observations (listing any manual writing and/or markers located on the roll).

### 3. Digitization challenges

The continuous recording of the data is a real challenge for the digitization. The companies that are digitizing 35mm films containing moving pictures, i.e., a long series of still images, with a standard known and fixed aspect ratio. The digitization machines have to be adapted for our purpose in two obvious ways: the digitization set up must include a significant overlap between 2 successive snapshots (about 10% of their length); the snapshots must also be "over-scanned" in the direction across the film, in order to digitize annotations and temporal markers. For those two points, it is possible to adjust the existing scanning machines rather easily. Another challenging issue is the geometry of the optical setup. Our data is very sensitive to trapezoidal aberration (i.e. when one side of an image is larger than the other). The trapezoidal aberration is not critical of classical movie scan, as each successive image is scanned with the same aberration, which is barely noticeable if kept small. In our application, a change of a few pixels of length from one end of a single snapshot to the other will prevent us from directly superimposing the successive snapshots during the reconstruction phase. We will need to correct for this aberration and this will result in a blurring of the images due to the resampling process. A last (but not least) challenge is the intensity dynamics of the digitization. The devices used by the industry is limited to 10 bits recording (e.g., $2^{10}$ = 1024 levels of gray) of dynamical range. This should be sufficient for our application, if the black and white levels are correctly setup.

The digitizing companies are providing data in movie industry standards: either DPX of TIFF formats. They both contain full resolution uncompressed single images. DPX images can be converted into TIFF with usual image processing software (such as image-magick [4] on Linux). Standard TIFF processing tools are often using 8 or 16 bits of dynamics, so that all files have to be converted (from 10 bits to 16 bits).

The proposed image resolution for digitization is the HD resolution (1920x1080 pixels), with 10 bits per pixel. This results in ~20 MB single files, and each roll is fully scanned with about 1600 snapshots. This results in ~31 GB of disk space for a single film roll, and in ~45 TB for the full collection. The reconstruction of the scientific data will double this data volume. All raw image scans will be archived at the Bibliothèque of Observatoire de Paris (the library department). A collection of quick-look images (JPEG format) will be derived from the raw images, for easier display (e.g., on web pages). The reconstructed scientific datasets will be recorded in a science ready format (such as CDF).

## 4. Future steps

We expect to have a first series of digitized rolls by the end of 2018. The assessment of the data quality has to be set up. Concerning the reconstruction algorithm, preliminary testing has been conducted on a series of digitized samples. This first test allowed us to prepare the invitation to tender, reducing the risks. However, we plan to improve the algorithm and test new libraries, such as the photographic stitching software that are commonly used.

The rehabilitation step (transposing digitized raw data into scientific observations accessible to the community at standard format with associated documentation) will require a significant work from the NDA team, which will be carried on in the frame of national observation services activities.